%
%
%

\long\def\UN#1{$\underline{{\vphantom{\hbox{#1}}}\smash{\hbox{#1}}}$}
\def\NP{\vfil\eject}
\def\NI{\noindent}

\magnification=\magstep 2
\overfullrule=0pt
\hfuzz=16pt
\voffset=0.0 true in
\vsize=8.8 true in
\baselineskip 20pt
\parskip 6pt
\hoffset=0.1 true in
\hsize=6.3 true in
\nopagenumbers
\pageno=1
\footline={\hfil -- {\folio} -- \hfil}
 
\hfill{\tt Physics Letters A 226, 253-256 (1997)}
 
\hphantom{A} 

\centerline{\UN{\bf A Hamiltonian for Quantum Copying}}
 
\vskip 0.4in
 
\centerline{{\bf Dima Mozyrsky,}\ \ {\bf Vladimir Privman}}
 
\centerline{\sl Department of Physics, Clarkson University,
Potsdam, NY 13699, USA}
 
\centerline{{\bf and}}

\centerline{{\bf Mark Hillery}}
 
\centerline{\sl Department of Physics and Astronomy}
\centerline{\sl Hunter College of
the City University of New York}
\centerline{\sl 695 Park Avenue, New York, NY 10021,
USA}
 
\vfill
 
\centerline{\bf ABSTRACT}
 
We derive an explicit Hamiltonian for copying the basis up and down
states of a quantum two-state system---a qubit---onto $n$ \ ``copy''
qubits ($n\geq 1$) initially all prepared in the down
state. In terms of spin components, for spin-$1\over 2$ particle
spin states, the resulting Hamiltonian involves $n$- and $(n+1)$-spin
interactions. The case $n=1$ also corresponds to a quantum-computing
controlled-NOT gate.

\vfill
 
\NI {\bf PACS numbers:}\ 03.65.Bz, 85.30.St.
 
\NP

Interest in quantum computing [1-27] has boosted studies of
quantum mechanics of two-state systems such as the spin
states of spin-$1\over 2$ particles. We will use 
``spin'' to indicate a two-state system in this context.
The ``binary'' up and down spin states are of particular
significance and the two-state systems are also termed ``qubits''
in these studies.
While macroscopic ``desktop'' coherent quantum computational units
are still in the realm of science fiction [16,18],
miniaturization of computer components calls for consideration
of quantum-mechanical [14-16,18] aspects of
their operation. Experiments have recently been
reported [25,28-29] realizing the simplest quantum gates.
Decoherence effects [16,18,26,30-32]
and inherently quantum-mechanical computational
algorithms [30-31,33-36] have been studied.

Here we consider the signal-copying process in two-state systems.
Quantum copying is of interest also in cryptography 
and signal transmission [37-52]. The latter applications,
in their coherent-quantum-mechanical version, are on the
verge of being experimentally realized [38,39,41-43,47,48].  
We assume that $n+1$ spins are involved, where spin 1 is the input
which is prepared in the up state, $|1\rangle$, or down state,
$|0\rangle$, at time $t$. The aim is to obtain the same state
in the $n$ \ ``copy'' spin states, i.e., for spins $2,3,\ldots,n+1$, as
well as keep the original state of spin 1.
Generally, one cannot copy an arbitrary [53-56]
quantum state; however, one can duplicate a set of basis states such
as the qubit states considered here.
One can also discuss an approximate, optimized
copying of the linear combinations of the basis states [55,56].

Another limitation of the copying procedure [53-56] has been that the
{\it initial\/} state of the $n$ copy spins must be {\it fixed}. An
attempt to allow for a more general state leads to incomplete copying
which is also of interest [57]. In this work we assume that the initial
state, at time $t$, of all the copy spins is down, $|0\rangle$.
Our aim is to derive an explicit Hamiltonian for the copying process.

We adopt the approach in the quantum-computing
literature [1-27]
of assuming that a constant
Hamiltonian $H$ acts during the time 
interval $\Delta t$, i.e., we only consider evolution from $t$ to 
$t+\Delta t$. The dynamics can be externally timed, with $H$ 
being switched on at $t$ and off
at $t+\Delta t$. The time interval is then related to the 
strength of couplings in $H$ which are of order $\hbar/\Delta t$. 
Some time dependence can be allowed [27], of the form $f(t)H$, where
$f(t)$ averages to 1 over $\Delta t$ and vanishes outside this time
interval.

We will denote the qubit states by quantum numbers $q_j=0$ (down)
and $q_j=1$ (up), for spin
$j$. The states of the $n+1$ spins will then be expanded in the basis
$|q_1 q_2 \cdots q_{n+1}\rangle$. The actual copying process only
imposes the two conditions

$$ |1 0 0 \cdots 0 \rangle \to |1 1 1 \cdots 1 \rangle \;\; , \eqno(1)
$$
$$ |0 0 0 \cdots 0 \rangle \to |0 0 0 \cdots 0 \rangle \;\; , \eqno(2)
$$

\NI up to possible phase factors. Therefore, a unitary transformation
that corresponds to quantum evolution over the time interval
$\Delta t$ is by no mean unique (and so the Hamiltonian is not unique).
We will choose a particular transformation that allows analytical
calculation and, for $n=1$, yields a controlled-NOT Hamiltonian.
The controlled-NOT unitary transformations have been discussed in the
literature [7,13-15,28,58,59]. A recent preprint [59] also derives
an explicit Hamiltonian which is somewhat different from
ours; we compare and discuss both results later.  

We consider the following unitary transformation,

$$ \eqalign{U=&e^{i\beta}|111\cdots1\rangle \langle 100\cdots0| 
+e^{i\rho}|000\cdots0\rangle \langle 000\cdots0| \cr
+&e^{i\alpha}|100\cdots0\rangle \langle 111\cdots1|
+\sum\limits_{\left\{q_j\right\}'}
| q_1 q_2 q_3 \cdots q_{n+1} \rangle
\langle q_1 q_2 q_3 \cdots q_{n+1} | \; \; . } \eqno(3) $$

\NI Here the first two terms accomplish the desired copying
transformation. The third term is needed for unitarity (since the
quantum evolution is reversible). We allowed for general phase factors
in these terms. The sum in the fourth term, $\left\{q_j\right\}'$, is
over {\it all the other\/}
quantum states of the system, i.e., excluding the three
states $|111\cdots1\rangle$, $|100\cdots0\rangle$,
$|000\cdots0\rangle$. One could maintain analytical tractability while
adding phase factors for each term in this sum; however, the added
terms in the Hamiltonian are not interesting. One can check by explicit
calculation that $U$ is unitary, $U^\dagger U=1$.

To calculate the Hamiltonian $H$ according to

$$ U=e^{-iH\Delta t / \hbar} \;\; , \eqno(4) $$

\NI we diagonalize $U$. The
diagonalization is simple because we only have
to work in the subspace of the three special states identified in (3),
see the preceding paragraph. Furthermore, the part related to the state
$|000\cdots0\rangle$ is diagonal. In the subspace labeled by
$|111\cdots1\rangle$, $|100\cdots0\rangle$, $|000\cdots0\rangle$, in
that order, the operator $U$ is represented by the matrix

$$ {\cal U}=\pmatrix{ 0 & e^{i\beta} & 0 \cr
 e^{i\alpha} & 0 & 0 \cr
 0 & 0 & e^{i\rho} } \;\; . \eqno(5) $$

The eigenvalues of $\cal U$ are
$e^{i(\alpha+\beta)/2}$, $-e^{i(\alpha+\beta)/2}$, $e^{i\rho}$.
Therefore the eigenenergies of the Hamiltonian in the selected
subspace can be of the form 

$$ E_1=-{\hbar \over 2 \Delta t}(\alpha+ \beta)+{2\pi \hbar
\over \Delta t}N_1 \;\; , \eqno(6) $$

$$ E_2=-{\hbar \over 2 \Delta t}(\alpha+ \beta)+{2\pi \hbar
\over \Delta t}\left(N_2+{1\over 2}\right) \;\; , \eqno(7) $$

$$ E_3=-{\hbar \over \Delta t} \rho + {2\pi \hbar
\over \Delta t}N_3 \;\; , \eqno(8) $$

\NI where $N_{1,2,3}$ are arbitrary integers. 

In order to simplify the expressions, we will limit our consideration
to a particular set of parameters. We would like to minimize energy gaps
of the Hamiltonian [27] and generally, keep
the energy spectrum symmetric.
The latter condition yields a more elegant
answer; actually, analytical calculation is possible
with general parameter values.
Thus, we take $\rho=0$, $N_3=0$, and also impose the
condition $E_1+E_2=0$.
We then take the diagonal matrix with $E_{1,2,3}$ as diagonal elements
and apply the inverse of the unitary transformation that
diagonalizes $\cal U$. All the calculations are straightforward and
require no further explanation or presentation of details in the
matrix notation. We note, however, that one could do all these
calculations directly in the qubit-basis notation such as in (3);
the diagonalization procedure is then the 
Bogoliubov transformation familiar from solid-state physics.

The result for the Hamiltonian in the three-state subspace is the matrix

$$ {\cal H}={\pi\hbar \over \Delta t}\left(N-{1\over 2}\right)
\pmatrix{0&e^{-i\gamma}&0\cr
 e^{i\gamma}&0&0\cr
 0&0&0} \;\; , \eqno(9) $$

\NI which depends on one real parameter

$$ \gamma={\alpha+\beta \over 2} \;\; \eqno(10) $$

\NI and on one arbitrary integer

$$ N=N_1-N_2 \;\; . \eqno(11) $$

The full Hamiltonian $H$ in the $2^{n+1}$-dimensional spin space is

$$ H= {\pi\hbar \over \Delta t}\left(N-{1\over 2}\right)
\Big(e^{-i\gamma}|111\cdots1\rangle \langle 100\cdots0|
+e^{i\gamma}|100\cdots 0\rangle
\langle 111\cdots 1|\Big) \;\; . \eqno(12) $$

\NI In what follows we make the choice $N=1$ to simplify the notation.
The form of the Hamiltonian is misleading in
its simplicity. It actually involves $n$- and $(n+1)$-spin
interactions. To see this, we rewrite it
in terms of direct products of the unit matrices and
the standard Pauli matrices for spins $1, \ldots, n+1$,
where the spins are indicated by superscripts (and $N=1$):

$$ H={\pi\hbar \over 2^{n+2} \Delta t}\left(1+\sigma_z^{(1)}\right)
\left(e^{-i\gamma}
\sigma_+^{(2)} \sigma_+^{(3)} \cdots \sigma_+^{(n+1)} 
+e^{i\gamma} \sigma_-^{(2)} \sigma_-^{(3)} \cdots \sigma_-^{(n+1)}
\right) \;\; ; \eqno(13) $$

\NI here $\sigma_\pm=\sigma_x \pm i\sigma_y$; \ $\sigma_+=
\pmatrix{0&2 \cr 0&0 }$, \ $\sigma_-=
\pmatrix{0&0 \cr 2&0 }$.

Multispin interactions are much less familiar and
studied in solid-state and other systems than
two-spin interactions. Therefore, the fact that for $n=1$
only single- and two-spin interactions are present is significant.
In actual quantum-computing and other applications it may be more
practical to make copies in stages, generating only one copy in
each time interval, rather than produce $n>1$ copies simultaneously.
Let us explore the $n=1$ case further. The Hamiltonian (with $N=1$)
is, in terms of spin components (or rather the Pauli
matrices to which the spin-component operators are proportional),

$$ H_{n=1}={\pi\hbar \over 4 \Delta t}\left(1+\sigma_z^{(1)}\right)
\left[(\cos \gamma) \sigma_x^{(2)} + (\sin \gamma) \sigma_y^{(2)}\right]
\;\; . \eqno(14) $$

\NI This Hamiltonian involves two-spin couplings and also interactions
which are linear in the $x$ and $y$ spin components. The latter may be
due to a magnetic field applied in the $xy$-plane, at an angle $\gamma$
with the $x$ axis.

Finally, we note that the $n=1$ ``single-copy'' Hamiltonian also
describes the controlled-NOT quantum gate with the same input and
output spins. The truth table for the classical controlled-NOT can
be written as follows in terms of the qubit states:

$$ |11\rangle \to |10\rangle \;\; , \eqno(15)$$
$$ |10\rangle \to |11\rangle \;\; , \eqno(16)$$
$$ |01\rangle \to |01\rangle \;\; , \eqno(17)$$
$$ |00\rangle \to |00\rangle \;\; . \eqno(18)$$

\NI The ``control'' spin, 1, being up causes the other spin, 2, to flip.
The control being down causes the second spin not to change.

The controlled-NOT unitary transformations have been discussed in the
literature [7,13-15,28,58,59]. It is obvious that in
the four-dimensional two-spin space labeled by
$|11\rangle$, $|10\rangle$, $|01\rangle$, $|00\rangle$, in that order,
the most general transformation matrix is of the form

$$ U=\pmatrix{0&e^{i\beta}&0&0\cr
 e^{i\alpha}&0&0&0\cr
 0&0&e^{i\omega}&0\cr
 0&0&0&e^{i\rho}} \;\; . \eqno(19)$$

\NI Our selected Hamiltonian accomplishes
such a transformation (for $n=1$
only). The matrix $U$ corresponding to (14) has the
following choice of the phase factors,

$$ U_{n=1}=\pmatrix{0&-ie^{-i\gamma}&0&0\cr
 -ie^{i\gamma}&0&0&0\cr
 0&0&1&0\cr
 0&0&0&1}\;\; . \eqno(20)$$

\NI Note that the details of this result depend on us setting $N=1$.

A recent preprint [59] presents another controlled-NOT Hamiltonian.
Their Hamiltonian corresponds to all phases zero is (19):
$\alpha=\beta=\omega=\rho=0$. In our notation, their Hamiltonian
corresponds to putting $\gamma=0$ in (14) and also adding a term
{\it linear\/} in $1+\sigma_z^{(1)}$. The latter addition only
affects the phases, and only in the upper-left quadrant of (19),
(20), and it can
be adjusted to yield a $U$ matrix with all nonzero elements equal 1
which is perhaps aesthetically more appealing than (20).
The following Hamiltonians (there are infinite number of possible
ones) are the simplest in this family:

$$ H_{\rm controlled\hbox{-}NOT}=\pm{\pi\hbar
\over 4 \Delta t}\left(1+\sigma_z^{(1)}\right)
\left(1-\sigma_x^{(2)}\right)
\;\; . \eqno(21) $$

In summary, we derived explicit Hamiltonians for $n$-copy
quantum copying.
For $n=1$, the interactions are the most useful because
they involve at most two-spin couplings. Furthermore, the $n=1$
Hamiltonian also
corresponds to the controlled-NOT gate.

This work has been supported in part by a US Air
Force grant, contract number F30602-96-1-0276. 
This financial assistance is gratefully acknowledged.

\NP
 
\centerline{\bf REFERENCES}{\frenchspacing
 
\item{[1]} A. Barenco, 
Proc. R. Soc. Lond. A {\bf 449}, 679 (1995).

\item{[2]} A. Barenco, ``Quantum 
Physics and Computers'' (preprint).

\item{[3]} A. Barenco, C.H. Bennett, 
R. Cleve, D.P. DiVincenzo,
N. Margolus, P. Shor, T. Sleator, J.A. Smolin and H. Weinfurter,
Phys. Rev. A {\bf 52}, 3457 (1995).

\item{[4]} G. Brassard, ``New Trends 
in Quantum Computing'' (preprint).

\item{[5]} P. Benioff, J. Stat. Phys. {\bf 29}, 515 (1982).

\item{[6]} C.H. Bennett,
Physics Today, October 1995, p. 24.

\item{[7]} J.I. Cirac and P. Zoller, 
Phys. Rev. Lett. {\bf 74}, 4091 (1995).

\item{[8]} D. Deutsch, Physics World, June 1992, p. 57.

\item{[9]} D. Deutsch, A. Barenco and A. Ekert, 
Proc. R. Soc. Lond. A {\bf 449}, 669 (1995).

\item{[10]} D.P. DiVincenzo, Science {\bf 270}, 255 (1995). 

\item{[11]} D.P. DiVincenzo, 
Phys. Rev. A {\bf 51}, 1015 (1995).

\item{[12]} A. Ekert, ``Quantum Computation'' (preprint). 

\item{[13]} A. Ekert and R. Jozsa, Rev. Mod. 
Phys. (to appear).

\item{[14]} R. Feynman, Int. J. Theor. 
Phys. {\bf 21}, 467 (1982).

\item{[15]} R. Feynman, Optics News 
{\bf 11}, 11 (1985).

\item{[16]} S. Haroche and J.-M. Raimond, 
Physics Today, August 1996, p. 51.

\item{[17]} S.P. Hotaling, ``Radix-$R>2$ 
Quantum Computation'' (preprint).

\item{[18]} R. Landauer, 
{Philos. Trans. R. Soc. London Ser.} 
A {\bf 353}, 367 (1995). 

\item{[19]} S. Lloyd, Science {\bf 261}, 1563 (1993). 

\item{[20]} N. Margolus, ``Parallel Quantum 
Computation'' (preprint).

\item{[21]} A. Peres, Phys. Rev. A {\bf 32}, 3266 (1985). 

\item{[22]} D.R. Simon, ``On the Power of 
Quantum Computation'' (preprint). 

\item{[23]} A. Steane, ``The Ion Trap 
Quantum Information Processor'' (preprint). 

\item{[24]} B. Schumacher, Phys. Rev. A {\bf 51}, 2738 (1995). 

\item{[25]} B. Schwarzschild, Physics Today, March 1996, p. 21. 

\item{[26]} W.H. Zurek, Phys. Rev. Lett. {\bf 53}, 391 (1984). 

\item{[27]} D. Mozyrsky, V. Privman and S.P. Hotaling,
``Design of Gates for Quantum Computation: the NOT Gate'' (preprint).

\item{[28]} C. Monroe, D.M. Meekhof, B.E. King, 
W.M. Itano and D.J. Wineland, 
Phys. Rev. Lett. {\bf 75}, 4714 (1995).

\item{[29]} Q. Turchette, C. Hood, 
W. Lange, H. Mabushi and H.J. Kimble,
Phys. Rev. Lett. {\bf 75}, 4710 (1995). 

\item{[30]} I.L. Chuang, R. Laflamme, 
P.W. Shor and W.H. Zurek,
Science {\bf 270}, 1633 (1995).

\item{[31]} E. Knill and R. Laflamme,
``A Theory of Quantum 
Error-Correcting Codes'' (preprint). 

\item{[32]} W.G. Unruh, Phys. Rev. A {\bf 51}, 992 (1995). 

\item{[33]} C. D\"urr and P. H\o yer, 
``A Quantum Algorithm for Finding the Minimum'' (preprint).

\item{[34]} R.B. Griffiths and C.-S. Niu, 
``Semiclassical Fourier Transform for Quantum
Computation'' (preprint).

\item{[35]} L.K. Grover, ``A Fast Quantum 
Mechanical Algorithm for Estimating the Median'' (preprint).

\item{[36]} P.W. Shor, ``Algorithms 
for Quantum Computation: Discrete Log and 
Factoring. Extended Abstract'' (preprint). 

\item{[37]} A. Ekert, Nature {\bf 358}, 14 (1992).

\item{[38]} C.H. Bennett, G. Brassard and A. Ekert, Scientific
American, October 1992, p. 26.

\item{[39]} C.H. Bennett, F. Bessette, G. Brassard, L. Savail and J.
Smolin, J. Cryptology {\bf 5}, 3 (1992).

\item{[40]} C.H. Bennett, G. Brassard, C. Cr\' epeau, R. Jozsa, A.
Peres and W.K. Wooters, Phys. Rev. Lett. {\bf 70}, 1895 (1993).

\item{[41]} A. Muller, J. Breguet and N. Gisin, Europhys. Lett. {\bf
23}, 383 (1993).

\item{[42]} P.D. Townsend, J.G. Rarity and P.R. Tapster, Electron.
Lett. {\bf 29}, 1291 (1993).

\item{[43]} P.D. Townsend, Electron. Lett. {\bf 30}, 809 (1993).

\item{[44]} A.K. Ekert, B. Huttner, M.G. Palma and A. Peres,
Phys. Rev. A {\bf 50}, 1047 (1994).

\item{[45]} B. Huttner and A. Peres, J. Mod. Opt. {\bf 41}, 2397
(1994).

\item{[46]} B. Huttner and A.K. Ekert, J. Mod. Opt. {\bf 41}, 2455
(1995).

\item{[47]} J.D. Franson and B.C. Jacobs, Electron. Lett. {\bf 31},
232 (1995).

\item{[48]} A. Muller, H. Zbinden and N. Gisin, Europhys. Lett.
{\bf 33}, 335 (1996).

\item{[49]} J.I. Cirac, P. Zoller, H.J. Kimble and H. Mabuchi,
``Quantum State Transfer and Entanglement Distribution Among
Distant Nodes in a Quantum Network'' (preprint).

\item{[50]} B. Huttner, N. Imoto, N. Gisin and T. Mor, ``Quantum
Cryptography with Coherent States'' (preprint).

\item{[51]} A. Peres, ``Unitary Dynamics for Quantum Codewords''
(preprint).

\item{[52]} T. Mor, ``Reducing Quantum Errors and Improving Large Scale
Quantum Cryptography'' (preprint).

\item{[53]} W.K. Wooters and W.H. Zurek, Nature {\bf 299}, 802
(1982).

\item{[54]} D. Dieks, Phys. Lett. {\bf 92} A, 271
(1982).

\item{[55]} V. Bu\v zek and M. Hillery, ``Quantum Copying: Beyond the
No-Cloning Theorem'' (preprint).

\item{[56]} V. Bu\v zek and M. Hillery, in preparation.

\item{[57]} D. Mozyrsky and V. Privman, ``Quantum signal splitting as
entanglement due to three-spin interactions'' (preprint).

\item{[58]} C.H. Bennett, G. Brassard, S. Popescu, B. Schumacher,
J.A. Smolin and W.K. Wooters, ``Purification of Noisy Entanglement
and Faithful Teleportation via Noisy Channels'' (preprint).

\item{[59]} I.L. Chaung and Y. Yamamoto, ``The Persistent Qubit''
(preprint).

}

\bye